\documentclass{PoS}

\ShortTitle{Renormalization Group approach to Gravity}

\title{Renormalization Group approach to Gravity: the running of $G$ and $\Lambda$ inside galaxies and additional details on the elliptical  NGC 4494}

\author{\speaker{Davi C. Rodrigues}$^a$ , Oliver F. Piattella,$^a$ J\'ulio C. Fabris$^a$, Ilya L. Shapiro$^b$ \\
$^a$ Departamento de F\'{\i}sica, Universidade Federal do Esp\'{\i}rito Santo,  29075-910, Vit\'oria, ES, Brazil \\ $^b$Departamento de F\'{\i}sica, Universidade Federal de Juiz de Fora,  36036-330, Juiz de Fora, MG, Brazil \\
E-mail: \email{davi.rodrigues@ufes.br}}

\abstract{We explore the phenomenology of nontrivial quantum effects on low-energy gravity. These effects come from the running of the gravitational coupling parameter $G$ and the cosmological constant $\Lambda$ in the Einstein-Hilbert action, as induced by the Renormalization Group (RG). The Renormalization Group corrected General Relativity (RGGR model) is used to parametrize these quantum effects, and it is assumed that the dominant dark matter-like effects inside galaxies is due to these nontrivial RG effects. Here we present additional details on the RGGR model application, in particular on the Poisson equation extension that defines the effective potential, also we re-analyse the ordinary elliptical galaxy NGC 4494 using a slightly different model for its baryonic contribution, and  explicit solutions are presented for the running of  $G$ and $\Lambda$. The values of the NGC 4494 parameters as shown here have a better agreement with the general RGGR picture for galaxies, and suggest a larger radial 
anisotropy than the previously published result.}

\FullConference{VIII International Workshop on the Dark Side of the Universe,\\
		June 10-15, 2012\\
		Rio de Janeiro, Brazil}

\begin{document}

\renewcommand{\vec}[1]{{\bf #1}}
\renewcommand{\Re}{\,\mbox{Re}\,}
\renewcommand{\Im}{\,\mbox{Im}\,}

\def\<{\left \langle}
\def\>{\right \rangle}
\def\[{\left\lbrack}
\def\]{\right\rbrack}
\def\({\left(}
\def\){\right)}
\newcommand{\be}{\begin{equation}}
\newcommand{\ee}{\end{equation}}
\newcommand{\ea}{\end{eqnarray}}
\newcommand{\ba}{\begin{eqnarray}}
\newcommand{\prt}{{\partial}}
\newcommand{\diag}{\mbox{diag}}
\newcommand{\tr}{\mbox{tr}}
\newcommand{\grad}{\ensuremath{\vec{\nabla}}}
\newcommand{\bs}{\begin{sideways}}
\newcommand{\es}{\end{sideways}}
\newcommand{\chir}{{\chi^2_{\mbox{\tiny{red}}}}} 
\newcommand{\Newt}{N}
\newcommand{\Iso}{{\mbox{\tiny Iso}}}
\newcommand{\mond}{{\mbox{\tiny MOND}}}
\newcommand{\stvg}{{\mbox{\tiny STVG}}}
\newcommand{\IsoInf}{{\mbox{\tiny Iso} \infty}}
\newcommand{\dg}{\dagger} 
\newcommand{\ML}{$\Upsilon_*^B/ (\frac{M_\odot}{L_{\odot,\mbox{\tiny B}}})$}
\newcommand{\amond}{$\frac{a_0}{1.35\times10^{-8} \mbox{\tiny cm/s}^2}$}
%Only useful for internal organization of the tables.
\newcommand{\mnras}{Mon. Not. R. Astron. Soc.}
\newcommand{\aap}{Astronomy $\&$ Astrophysics}
\newcommand{\apjs}{ApJS}
\newcommand{\apjl}{Astrophys. J. Letters}
\newcommand{\apj}{Astronphys. J.}
\newcommand{\aj}{Astron. J.}
\newcommand{\pasa}{PASA}
\newcommand{\RGGR}{{\mbox{\tiny RGGR}}}
\newcommand{\MOND}{{\mbox{\tiny MOND}}}
\newcommand{\Ser}{{\mbox{\tiny S}}}
\newcommand{\ext}{{\mbox{\tiny ext}}}

\def\beq{\begin{eqnarray}}
\def\eeq{\end{eqnarray}}
\def\ln{\,\mbox{ln}\,}
\def\Det{\,\mbox{Det}\,}
\def\det{\,\mbox{det}\,}
\def\tr{\,\mbox{tr}\,}
\def\diag{\,\mbox{diag}\,}
\def\Tr{\,\mbox{Tr}\,}
\def\sTr{\,\mbox{sTr}\,}
\def\Res{\,\mbox{Res}\,}

\def\lap{\Delta}
\def\sla{\!\!\!\slash}
\def\al{\alpha}
\def\bet{\beta}
\def\ch{\chi}
\def\ga{\gamma}
\def\de{\delta}
\def\vp{\varepsilon}
\def\ep{\epsilon}
\def\ze{\zeta}
\def\io{\iota}
\def\ka{\kappa}
\def\la{\lambda}
\def\na{\nabla}
\def\pa{\partial}
\def\ro{\varrho}
\def\rh{\rho}
\def\si{\sigma}
\def\om{\omega}
\def\ph{\varphi}
\def\ta{\tau}
\def\th{\theta}
\def\te{\vartheta}
\def\up{\upsilon}
\def\Ga{\Gamma}
\def\De{\Delta}
\def\La{\Lambda}
\def\Si{\Sigma}
\def\Om{\Omega}
\def\Te{\Theta}
\def\Th{\Theta}
\def\Up{\Upsilon}
%%%%%%%%%%%%%%%%%%%%%%%%%%%%%%%%%%%%%%%%%%%%%%%%%%%%%%%%%%%%%%

%%%%%%%%%%%%%%%%%%%%%%%%%%%%%%%%%%%%%%%%%%%%%%%%%%%%%%%%%%%%%%
\section{Introduction} \label{intro}

In Refs. \cite{Rodrigues:2009vf, Rodrigues:2012qm, Rodrigues:2011cq, Farina:2011me, Fabris:2012wg, Rodrigues:2012wk} we presented new results on the application of renormalization group (RG) corrections to General Relativity in the astrophysical domain, in particular on a possible relation between RG large scale effects and dark matter-like effects in galaxies. The resulting phenomenological model was named RGGR. These developments were directly based on the RG application to gravity of Ref. \cite{Shapiro:2004ch}, and are consistent with the phenomenological consequences of diverse approaches to the subject, including the related to the asymptotic safety scenario of quantum gravity, see in particular Refs.  \cite{Reuter:2004nx,Reuter:2004nv,Reuter:2007de}.

Currently, in the context of quantum field theory in curved space time, it is impossible to construct a formal proof that the coupling parameter $G$ is a running parameter in the infrared. However, this possibility can not be ruled out.  The possibility of General Relativity being modified in the far infrared due to the renormalization group (RG) has been considered in different contexts, for instance,  \cite{Goldman:1992qs, Bertolami:1993mh, Dalvit:1994gf, Bertolami:1995rt}. The previous attempts to apply this picture to galaxies have considered for simplicity point-like galaxies (e.g., \cite{Shapiro:2004ch, Reuter:2004nx}). We extended previous considerations by identifying a proper renormalization group energy scale $\mu$  and by evaluating the consequences considering the observational data of disk \cite{Rodrigues:2009vf} and elliptical \cite{Rodrigues:2012qm} galaxies. We proposed in Ref. \cite{Rodrigues:2009vf} the existence of a relation between $\mu$ and the local value of the Newtonian  potential (
this relation was reinforced  afterwards \cite{Domazet:2010bk}). With this choice, the renormalization group-based approach (RGGR) was capable to mimic dark matter effects with great precision. Also, it is remarkable that this picture induces a very small variation on the gravitational coupling parameter $G$, namely a variation of about $10^{-7}$ of its value across a galaxy (depending on the matter distribution). We call this model RGGR, in reference to Renormalization Group corrected General Relativity.

\section{The RGGR effective potential and the running of $G$ and $\Lambda$}

This section is, in part, a review on the RGGR dynamics, but the route to deduce the effective potential is new, and an explicit expression for the modified Poisson equation (for spherical symmetry) is here presented. Additional details on this deduction will appear in Ref. \cite{RGGRspherical}. Also, from this route comes the equation that governs the $\Lambda$ running inside galaxies, which was used in Ref. \cite{Rodrigues:2012qm}. Another approach to find the effective potential, which was used previously, is to use a conformal transformation. 

\subsection{Effective potential and the modified Poisson equation}

The gravitational coupling parameter $G$ may  behave as a true constant in the far  IR limit, leading to standard General Relativity in such limit. Nevertheless, in the context of QFT in curved space time, there is no proof on that. According to Refs. \cite{Shapiro:2004ch, Farina:2011me}, a certain logarithmic running of $G$  is a direct consequence of covariance and must hold in all loop orders.  Hence the situation is as follows: either there is no new gravitational effect induced by the renormalization group in the far infrared, or there are such deviations and the gravitational coupling runs as

\be
	 \beta_{G^{-1}} \equiv \mu \frac{dG^{-1}}{d \mu} = 2 \nu \,  \frac{M_{\mbox{\tiny Planck}}^{2}}{c \, \hbar} = 2 \nu G_0^{-1}.
	\label{betaG}
\ee
Equation (\ref{betaG}) leads to the logarithmically  varying $G(\mu)$ function,
\be
	\label{gmu}
	G(\mu) = \frac {G_0}{ 1 + \nu \ln(\mu^2/\mu_0^2)},
\ee
where $\mu_0$ is a reference scale introduced such that $G(\mu_0) =G_0 $. The constant $G_0$ is the gravitational constant as measured in the Solar System. Actually, there is no need to be very precise on where $G$ assumes the value of $G_0$, due to the smallness of the variation of $G$; moreover, within the weak field limit, the value of $\mu_0$ is dynamically irrelevant, as it will be shown. The dimensionless constant $\nu$ is a phenomenological parameter which depends on the details of the quantum theory leading to eq. (\ref{gmu}). Since we have no means to compute the latter from first principles, its value should be fixed from observations.  The first possibility, namely of no new gravitational effects in the far infrared, corresponds to $\nu=0$.

The form of the action in question is simply the Einstein-Hilbert one, in which $G$ appears inside the integral, namely,\footnote{We use the $(- + + +)$ space-time signature.}
\be
	S_{\mbox{\tiny RGGR}}[g] = \frac {c^3}{16 \pi }\int \frac {R  - 2 \Lambda } G \, \sqrt{-g} \,  d^4x.
	\label{rggraction}
\ee
In the above, $G$ and $\Lambda$ are external scalar fields, that is, their dynamics do not come from pure classical arguments, but are given from the RG equations; in particular $G$ satisfies (\ref{gmu}).  For the problem of internal galaxy kinematics,  the cosmological constant effects are negligible (as expected and as shown in \cite{Rodrigues:2012qm}). 

The above, together with some  energy-momentum tensor $T_{\mu \nu}$, leads to the following field equations (see, e.g. \cite{Reuter:2003ca,Rodrigues:2012qm})
\be
	 G_{\mu \nu} +  \Lambda \,  g_{\mu \nu} + G \, \square G^{-1} g_{\mu \nu} - G \, \nabla_\mu \nabla_\nu G^{-1} = \frac{8 \pi G}{c^4} \, T_{\mu \nu} \, .
	\label{rggreq}
\ee

To find solutions for static spherically symmetric space-times (a natural symmetry for some ellipticals, including NGC 4494), the following line element will be considered,
\be
	ds^2 = - e^{\kappa(r)} c^2 dt^2 +  e^{\eta(r)} dr^2 + r^2 d\theta^2 + r^2 \sin(\theta) d\phi^2.
	\label{lineelement}
\ee
To be clear, by weak field it is mean that both $\kappa$ and $\eta$ are much smaller than one. Also it will be assumed ``small RG corrections'', i.e. that $G(\mu)$, as given by (\ref{gmu}), satisfies $|G(\mu)/G_0 -1|\ll 1$. 

From the geodesics, one finds that $\kappa$ acts as the effective potential, in the sense that its gradient is proportional to the acceleration felt by a test particle. But contrary to General Relativity, $\kappa$ does not satisfy a Poisson equation. By solving the field equations in the presence of ``dust'' of density $\rho(r)$,
\be
	\nabla^2 \[ \kappa + 2 \nu \ln \( \frac{\mu}{\mu_0}\) \] =  \frac{8 \pi G_0}{c^2} \rho.
\ee
Therefore the relation between the effective potential $\Phi$ to the Newtonian one $\Phi_N$ can be stated as 
\be
	  \Phi  =  \Phi_N -  2 \nu c^2 \ln \( \frac{\mu}{\mu_1}\),
	\label{kappaPhiN}
\ee
where $\mu_1$ is an arbitrary constant (it needs not to be $\mu_0$) and $\Phi \equiv \kappa c^2 / 2$.

\subsection{The energy scale setting}

In order to derive a test particle acceleration, we have to specify the proper energy scale $\mu$ for the problem setting in question, which is a time-independent gravitational phenomena in the weak field limit. This is a recent area of exploration of the renormalization group application, where the usual procedures for high energy scattering of particles cannot be applied straightforwardly. In \cite{Rodrigues:2009vf} we introduced a $\mu$ identification that seems better justified both from the theoretical and observational points of view. The characteristic weak-field gravitational energy scale does not comes from the geometric scaling $1/r$, but should be found from the Newtonian potential $\Phi_\Newt$, the latter is the field that characterises gravity in such limit. Therefore,
\be
	\frac{\mu}{\mu_0} = f\( \frac{\Phi_N}{\Phi_0}\).
	\label{muf}
\ee
If $f$ would be a complicated function with dependence on diverse constants, that would lead to a theory with small (or null)  prediction power. The simplest assumption, $ \mu \propto \Phi_\Newt$,  leads to $\mu \propto 1/r$ in the large $r$ limit; which is unsatisfactory on observational grounds (bad Newtonian limit and correspondence to the Tully-Fisher law). One way to recover the Newtonian limit is to impose a suitable cut-off, but this rough procedure does not solves the Tully-Fisher issues \cite{Shapiro:2004ch}. Another one is to use \cite{Rodrigues:2009vf}
\be
	\frac {\mu}{\mu_0} =\left( \frac{\Phi_\Newt}{\Phi_0} \right)^\alpha,
	\label{muphi}
\ee
where $\Phi_0$ and $\alpha$ are constants. The precise value of $\Phi_0$ is irrelevant for the dynamics in the weak field limit, since $\mu_0$ the dynamics in this limit does not depend on $\mu_0$ as argued above. The relevant parameter is $\alpha$. 

The parameter $\alpha$ is a phenomenological parameter that needs to depend on the mass of the system, and it must go to zero when the mass of the system goes to zero. This is necessary to have a good Newtonian limit. From the Tully-Fisher law, it is expected to increase monotonically with the increase of the mass of disk galaxies. For more details, see  \cite{Rodrigues:2012wk, Rodrigues:2012qm, FabrisFuture}.

\subsection{The running of $\Lambda$}

The running of $\Lambda$ can be stated as a differential equation that depends on $G$ and $R$ alone. From the energy-momentum conservation ($T_{\mu \nu}^{\; \; \; ;\nu} = 0$), together with the Bianchi identities ($G_{\mu\nu}^{\;\;\; ; \nu} = 0$) \cite{Rodrigues:2012qm},
\be
	\nabla_\nu \( \frac{\Lambda}G \) = \frac 12  R \, \nabla_\nu G^{-1},
	\label{EMrggr}
\ee
since $\(  \square \nabla_\nu - \nabla_\nu \square \) G^{-1} = R_{\nu \kappa} \nabla^\kappa G^{-1}$. The above result is exact. Equation (\ref{EMrggr})  also appeared in Refs. \cite{Koch:2010nn, Cai:2011kd}. And from eqs.(\ref{muphi}, \ref{EMrggr}), 
\be
	\Lambda'= 2 \nu \alpha \frac{\Phi_N'}{\Phi_N} \[ \frac{4 \pi G \rho}{c^2} +  \Lambda + 3 \nu \alpha \(\frac{\nabla^2 \Phi_N}{\Phi_N} - \frac{\Phi_N'^2}{\Phi_N^2}\)\].
	\label{lambdalinha}
\ee
Hence, from the knowledge of the matter distribution, one can find $\Phi_N$ and consequently $\Lambda(r)$.

\subsection{Application to spherically symmetric Jeans equation}
Since elliptical galaxies are mainly supported by velocity dispersions (VD), the main equation for galaxy  kinematics in this case is the following expression for the projected (line-of-sight) VD (see also \cite{Mamon:2004xk}),
\be
	 \sigma_p^2(R) = \frac {2 G_0}{I(R)} \int_R^\infty K\(\frac r R\) \frac{\ell(r) M(r) } r dr,
	 \label{sigma_pK}
\ee
where
{\small
\be
	 K(u) \equiv  \frac 12 u^{2 \beta - 1}\[ \(\frac 32 - \beta \) \sqrt \pi \frac{\Gamma(\beta - 1/2)}{\Gamma(\beta)} +  \beta B\(\frac 1{u^2}, \beta + \frac 12, \frac 12 \) -  B\(\frac 1{u^2}, \beta - \frac 12, \frac 12 \)\], \nonumber
\ee}
$B(x,a,b) = \int_0^x t^{a-1} (1-t^{b-1}) dt$ is the incomplete beta function, $\Gamma$ is the Gamma function, $r$ stands for the deprojected (spherical) radius, $R$ for the projected (line of sight) radius,  $I(R)$ is the luminosity intensity,  $\ell(r)$ is the luminosity density (found from the deprojection of $I(R)$), $\beta$ is the anisotropy parameter (it is zero if the galaxy has an isotropic VD profile) and $M(r)$ is the total (effective) mass of the system at the radius $r$. See Ref. \cite{Rodrigues:2012qm} and references therein for additional details. In the case of Newtonian gravity without dark matter, $M(r)$ would stand, within a good approximation, as the stellar mass inside the radius $r$, i.e., $M_*(r)$.

For RGGR without dark matter, the total mass inside the radius $r$ is  the baryonic mass $M_*(r)$. Nevertheless, the non-Newtonian gravitational effects of RGGR for spherically symmetric systems can be understood from the Newtonian perspective as if the total mass was given by the following total effective mass \cite{Rodrigues:2012qm}
 \be
	M(r) = M_*(r) + M_{\mbox{\tiny RGGR}}(r),
\ee
with 
\be
	M_{\mbox{\tiny RGGR}} (r) = \frac{\alpha \nu c^2}{G_0} \frac{r}{1 + \frac{4 \pi r }{M_*(r)} \int_{r}^{\infty} \rho_*(r') r' dr'}.
	\label{MRGGR}
\ee
The above is the the essential equation to model elliptical galaxies (compatible with the spherical approximation) within RGGR.

\section{NGC 4494 }

\subsection{Mass models}

NGC 4494 is an ordinary elliptical recently analysed within RGGR \cite{Rodrigues:2012qm}, and which has received considerable attention previously, in particular due to its apparent lack of dark matter \cite{Romanowsky:2003qv}.  Here we present a variation of the analysis presented in that Ref. \cite{Rodrigues:2012qm}, namely we do not directly use the photometric data with the S\'ersic extension to model the stellar mass, instead  only  the  S\'ersic  profile that best fit the surface brightness of this galaxy is used (an approach here labeled ``Full S\'ersic''). See Table \ref{N4494resultsTable} and Fig. \ref{N4494dsu}. 

In part this is relevant to evaluate to what extent  our results presented in \cite{Rodrigues:2012qm} are robust to small changes on the baryonic mass content. The main results are  i) within the 1 $\sigma$ error bars, the results are compatible; ii)  the best fit value for the anisotropy parameter considerably moves towards radial anisotropy (from about isotropy to $0.44$) (see further comments in the Conclusions).     

\begin{table*}[htdp] 
\begin{center}
{\footnotesize 
\begin{tabular}{l c c c c c c c}
\multicolumn{7}{c}{\emph{ \normalsize NGC 4494}}\\
\hline \hline
\multicolumn{7}{c}{\emph{RGGR without dark matter} } 				  \\ 
Stellar model $^{(1)}$				& $\alpha\,\nu \times 10^{7}$	& $\beta		$	&\ML 			& Mass ($10^{10} M_\odot$)		&${\chi^{2}}$	&${\chir}$\\ \cline{1-7}
Full S\'ersic+$\beta_{[0]}$ 			&$0.77\pm0.6$			&0				&$4.27\pm0.31$		&$10.1\pm1.5$ 				&9.02		&0.38\\
Full S\'ersic+$\beta_{[0,1]}$			&$2.6^{+0.99}_{-2.0}$		&$0.44^{+0.52}_{-0.41}$		&$3.4^{+0.98}_{-3.1}$	&$8.0^{+3.0}_{-7.9} $			&5.10		&0.22\\
Full S\'ersic+K.IMF+$\beta_{[0]}$		&$0.84\pm0.43$			&0				&$4.23\pm0.21$		&$10.0\pm1.3 $				&9.09		&0.36\\
Full S\'ersic+K.IMF+$\beta_{[0,1]}$ 		&$1.17\pm{0.70}$		&$0.19\pm0.19$			&$4.09\pm0.32$		&$9.7\pm 1.5$ 				&7.06		&0.29\\ 	

\hline \hline
\end{tabular}
\caption{\label{N4494resultsTable} \footnotesize NGC 4494 results for RGGR, assuming that the stellar mass profile is only given by a S\'ersic profile. This table extends a table on this galaxy in Ref.\cite{Rodrigues:2012qm}, see this reference for further details. (1) $\beta_{[0]}$ indicates isotropic VD, $\beta_{[0,1]}$ indicates constant anisotropy with $\beta \in [0,1]$ (except for the error bars, the above also holds for $\beta \in [-1,1]$), K.IMF is a reference to Kroupa IMF, and it means that the expected value of $\Upsilon_*$ was used in the $\chi^2$ definition (it was used $4.2 \pm 1.0 M_\odot / L_{\odot, B}$ \cite{2012ApJ...748....2D}). For details on this procedure see Ref. \cite{Rodrigues:2012qm}.  }}
\end{center}
\end{table*} 

\begin{figure*}[thbp]
\begin{center}
	  \includegraphics[width=85mm]{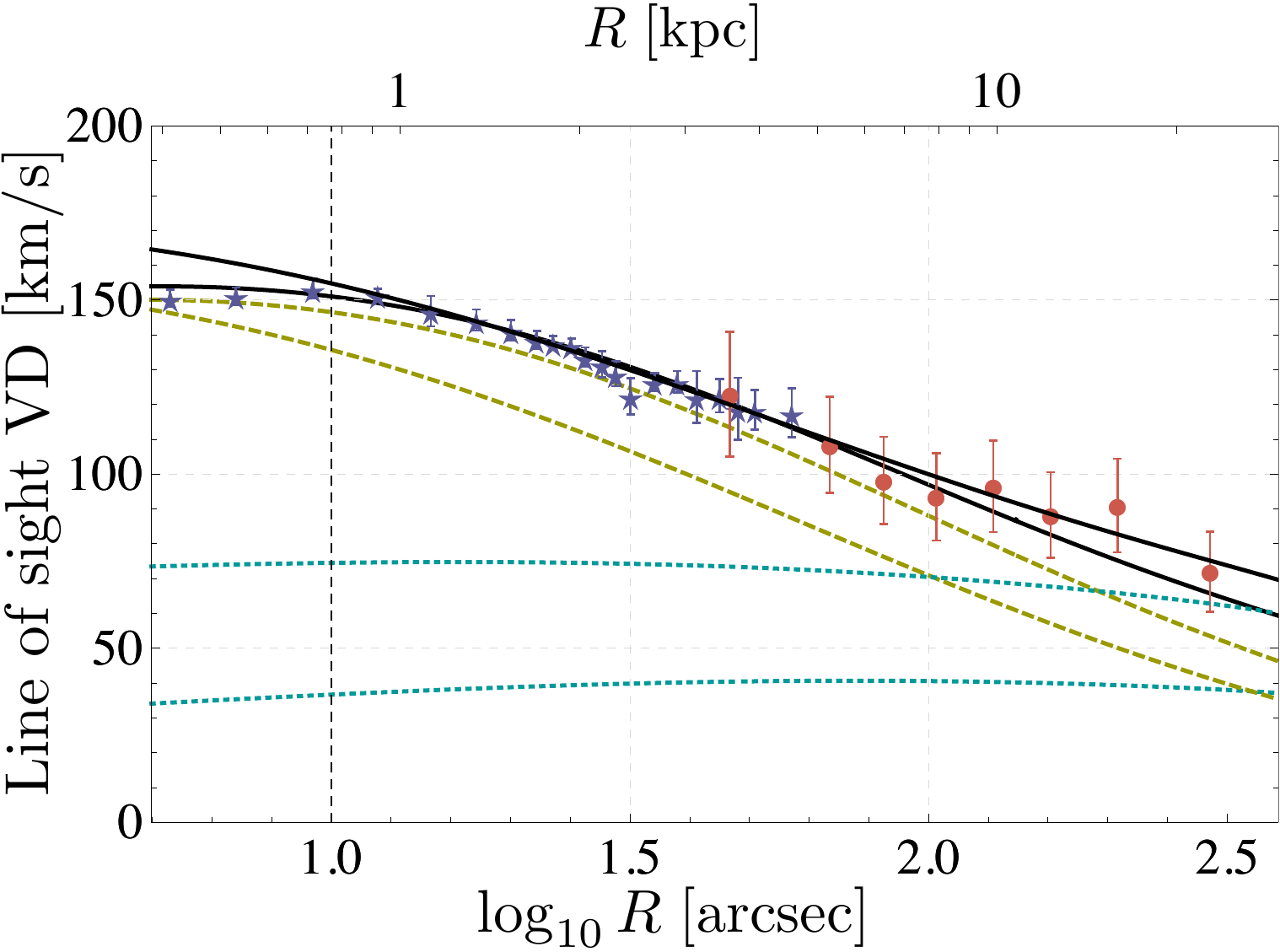}
\caption{Two NGC 4494 mass models within RGGR and without dark matter. The curves refer to mass models  composed by the stellar component (given by a S\'ersic profile) and RGGR gravity. The black solid lines are the resulting VD for each model, the yellow dashed and blue dotted lines are respectively the stellar Newtonian and non-Newtonian contributions to the total VD. One of the models assumes isotropy ($\beta = 0$),  the second   assumes $\beta \in [-1,1]$. The vertical dashed line signs the radius above which the observational data is considered  for the fitting procedure (10 arcsec).}
\label{N4494dsu}
\end{center}
\end{figure*}

The four models presented in Table \ref{N4494resultsTable} lead to total stellar masses that are inside the range expected from the Kroupa (and Chabrier) IMF's \cite{2012ApJ...748....2D}, namely  $M_* \in [8.0 M_\odot, 12.0 M_\odot]$. The found anisotropy values are also reasonable \cite{2012ApJ...748....2D}. On the other hand, only the second model has a value of $\alpha$ compatible with the baryonic matter amount of this galaxy\footnote{In accordance with expectations driven by the analysis of other galaxies, see the conclusions.}, all the others display a ``dearth of RG effects'' for the given baryonic content. This has some similarities with the $\Lambda$CDM case, where some radial anisotropy is needed to find reasonable amounts of dark matter \cite{Romanowsky:2003qv, 2009MNRAS.393..329N}.

\subsection{The running of $G$ and $\Lambda$ inside NGC 4494}

From the NGC 4494 results above, here the corresponding running of $G$ and $\Lambda$ inside this galaxy are shown. A priori, according to eqs. (\ref{gmu}, \ref{muphi}), $G(r)$ depends on the value of $\Phi_0$ (or $\mu_0$), nevertheless dynamically such constant plays no role (it can be anything and it will not have dynamical impact whenever the weak field approximation holds).   

\begin{figure*}[thbp]
\begin{center}
	  \includegraphics[width=150mm]{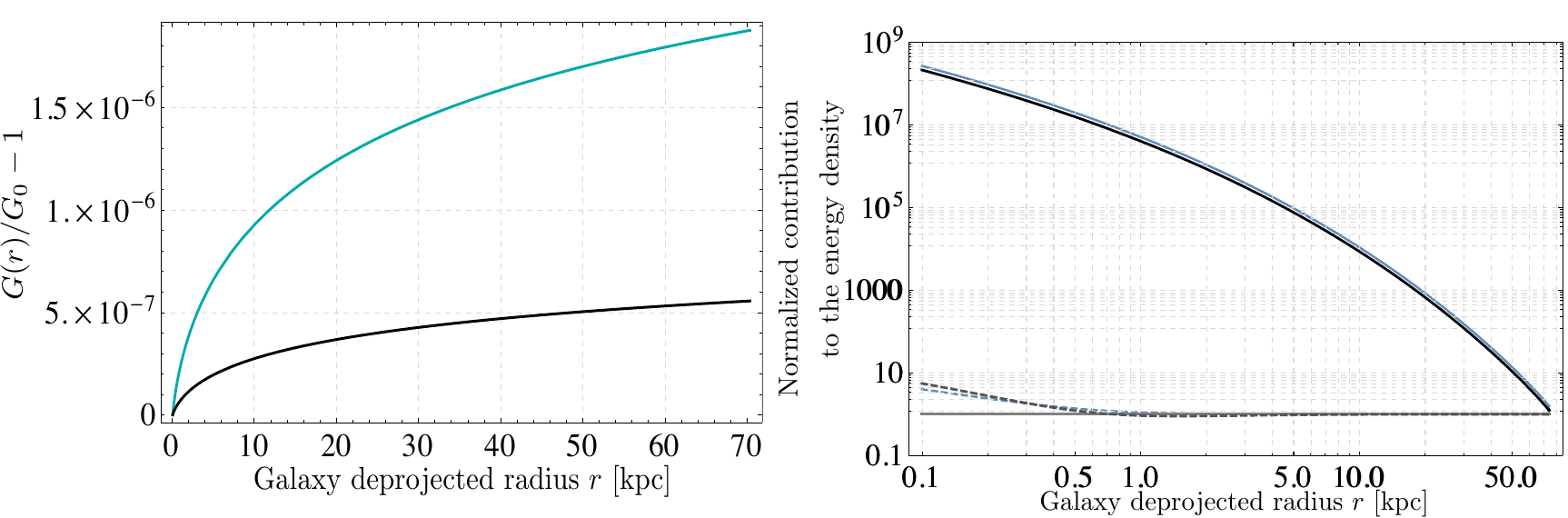}
\caption{The plots show the variation of $G$ (left) and $\Lambda$ (right) inside NGC 4494, both plots considers two models, ``Full S\'ersic+$\beta_{[0]}$'' (black on the left plot, blue on the right plot) and ``Full S\'ersic+$\beta_{[0,1]}$'' (cyan on the left plot, black on the right plot). Both of  the $G$ plots in the left are normalised with $G(0) = G_0$; the latter fixes the (dynamically irrelevant) constant $\Phi_0$ for each model. The right plot compares the ratio of the absolute value of $\Lambda(r) $ and $4 \pi G \rho_*/c^2$  to  $\Lambda(25 R_e)$ (for further details, see also \cite{Rodrigues:2012qm}).  The solid  curve corresponds to $4 \pi G\rho_*(r) /(c^2 \Lambda_0)$, the dashed  curve  to $\Lambda(r)/\Lambda_0$.  In both cases, the absolute value  of $\Lambda$, at any radius, is  too small to significantly change the average  internal galaxy dynamics, which is in accordance with the approximations used to derive the effective potential of RGGR.}
\label{LandGrunning}
\end{center}
\end{figure*}

\section{Conclusions}

In summary, RGGR is a model based on the theoretical possibility that the beta function of the gravitational coupling parameter $G$ may not be zero in the far infrared. Currently, there is no way to directly deduce this behaviour from first principles, nevertheless eq. (\ref{betaG}) is a natural, if not unique, possibility that has appeared many times before in the context of Quantum Field Theory in curved space-time.  %This equation depends on a universal free parameter $\nu$ that can be constrained from experiments and observations, with $\nu = 0$ corresponding to standard General Relativity. The eq. (\ref{betaG}) also depends on an energy scale $\mu$, which should be related to the symmetries and physical interactions that are being evaluated. Considering gravitation effects in stationary systems with weak Newtonian potential ($\Phi_N /c^2\ll 1$) and slow particle velocities ($v/c \ll 1$), it is natural to use a relation of the type (\ref{muf}). In Ref. \cite{Rodrigues:2009vf} the relation (\ref{muphi}) 

RGGR  without dark matter is a model with  one phenomenological free parameter ($\alpha$) which is capable of dealing with the kinematics of diverse galaxies. The $\alpha$  relation to other physical parameters is being disclosed in a work in progress \cite{FabrisFuture}. In particular, we are unveiling  a correlation between $\alpha$ and the galaxy baryonic mass. The galaxy that was in worst agreement with such correlation was NGC 4494, since it displayed a too low $\alpha \nu$ value for its mass (considering the solutions presented in Ref.\cite{Rodrigues:2012qm}). Here it is commented that with a small change on the modelling of the baryonic mass (i.e., modelling its mass exclusively from its S\'ersic profile, which is well inside the errors on the conversion from light to mass), a greater tendency towards higher radial anisotropy, lower mass and higher $\alpha$ was found (model with free constant anisotropy). The corresponding mass is still inside the mass range associated with the Kroupa IMF \cite{
2012ApJ...748....2D}, and fits much better within the general picture that is being unveiled \cite{FabrisFuture}.

\vspace{.2in}

\noindent
{\bf Acknowledgements } \\
DCR thanks the DSU organisers for the invitation and for the very nice meeting. DCR and JCF thank CNPq and FAPES for partial financial support.  The work of I.Sh. has been partially supported by CNPq, FAPEMIG and ICTP.

\bibliographystyle{JHEP}

%\bibliography{/Users/Davi/Desktop/Works/bibdavi2012}{} %Mac

\bibliography{/home/davi/Desktop/Works/bibdavi2012}{} %Linux

\end{document}